\documentstyle[12pt]{article}
\newcommand{\beq}{\begin{equation}}
\newcommand{\eeq}{\end{equation}}

\newcommand{\beqn}{\begin{eqnarray}}
\newcommand{\eeqn}{\end{eqnarray}}

\newcommand{\ba}{\bar \alpha}
\newcommand{\bb}{\bar \beta}

\begin{document}
\begin{center}
{\Large \bf Nucleon polarizability contribution
     to the hydrogen Lamb shift
     and hydrogen -- deuterium isotope shift \\}
\vspace*{1cm}

{\bf I.B. Khriplovich\footnote{ E-mail: khriplovich@inp.nsk.su}
 and R.A. Sen'kov\footnote{ E-mail: senkov@vxinp.inp.nsk.su} \\ }
{ Budker Institute of Nuclear Physics,\\}
{ 630090 Novosibirsk, Russia,\\}
{ and Novosibirsk University }
\end{center}

\vspace*{1.0cm}

\begin{abstract}
{ The correction to the hydrogen Lamb shift due to the proton
electric and magnetic polarizabilities is expressed analytically
through their static values, which are known from experiment. The
numerical value of the correction to the hydrogen 1S state
is $\;-\;71 \pm \,11 \pm\,7$ Hz.
Correction to the H-D 1S-2S -- isotope shift due to the proton and
neutron polarizabilities is estimated as $\;53 \pm \,9 \pm \, 11$ Hz.}
\end{abstract}

\bigskip

{\bf 1.}
High experimental precision attained in the hydrogen and deuterium
spectroscopy (see, e.g., \cite{han,err}) stimulates considerable
theoretical activity in this field. In particular, the deuteron
polarizability contribution to the Lamb shift in deuterium was
calculated in  \cite{pach}$-$\cite{fri1}. A special feature of 
these corrections is that they contain logarithm of
the ratio of a typical nuclear excitation energy to the electron mass,
$\ln \bar E/m_e$.

In the present note we consider the problem of the proton
polarizability correction to the Lamb shift in hydrogen. The typical
excitation energy for the proton $\bar E_p \sim 300$ MeV is large as
compared to other nuclei (to say nothing of the deuteron). So, 
$\ln \bar E_p/m_e$ is not just a mere theoretical parameter,
it is truly large, about $6 - 7$, which makes the logarithmic
approximation quite meaningful quantitatively.

In our calculation we follow closely the approach of \cite{khr}.
In particular, we use the gauge $A_0=0$ for virtual photons, so that
the only nonvanishing components of the photon propagator are
$ D_{im}=d_{im}/k^2,\;\; d_{im}=\delta_{im}- k_i k_m/{\omega}^2\;\;
(i,m=1,2,3)$. The electron-proton forward scattering amplitude, we are
interested in, is 
\begin{equation}\label{ep}
T=4\pi i \alpha \int \frac{d^4k}{(2\pi)^4}\;D_{im}D_{jn}
       \frac{\gamma_i({\hat l}-{\hat k}+m_e)\gamma_j}{k^2-2lk}\;M_{mn}.
\end{equation}
Here $l_{\mu}=(m_e,0,0,0)$ is the electron momentum. The nuclear-spin
independent Compton forward scattering amplitude, which is of
interest to us, can be written as
\begin{equation}\label{gp}
 M=\ba(\omega^2,{\bf k}^2){\bf E^*E}+\bb(\omega^2,{\bf k}^2){\bf B^*B}
 \;=M_{mn}e_me_n{}^*,
\end{equation}
where $\ba$ and $\bb$ are the nuclear electric and magnetic
polarizabilities, respectively.  The structure $ \gamma_i({\hat
l}-{\hat k}+m_e)\gamma_j$ in (\ref{ep}) reduces to $ -
\omega\delta_{ij} $.  Perhaps, the most convenient succession of
integrating expression (\ref{ep}) is as follows: the Wick
rotation; transforming the integral over the Euclidean $\omega$ to
the interval $(0,\,\infty)$; the substitution ${\bf k}\rightarrow
{\bf k}\,\omega$. Then the integration over $\omega$ is easily
performed with the logarithmic accuracy:
\begin{eqnarray}\label{c}
{\int_0}^{\infty}\frac{d\omega^2}{\omega^2+4{m_e}^2/(1+{\bf k}^2)^2}
  \;\left[(3+2{\bf k}^2+{\bf k}^4)\ba(-\omega^2,-\omega^2{\bf k}^2)-
          2{\bf k}^2\bb(-\omega^2,-\omega^2{\bf k}^2)
    \right]\;\\
    =\;\left[(3+2{\bf k}^2+{\bf k}^4)\ba(0)
                   -2{\bf k}^2\bb(0)\right]\;
    \ln\frac{\bar E ^2}{m_e^2}.\nonumber
\end{eqnarray}
The crucial point is that, within the logarithmic approximation, both
polarizabilities $\ba$ and $\bb$ in the lhs can be taken at
$\omega=0$, ${\bf k}^2=0$. The final integration over $d^3{\bf k}$
is trivial.

The resulting effective operator of the electron-proton interaction
(equal to $-T$) can be written in the coordinate representation as
\begin{equation}\label{d}
V=-\,\alpha m_e\, [\,5\ba(0)-\bb(0)\,]\;\ln\frac{\bar E}{m_e}\;
                                                       \delta({\bf r}).
\end{equation}
This expression applies within the logarithmic accuracy for arbitrary
nuclei. It should be mentioned that a similar relation for hydrogen
was obtained in \cite{star}, our numerical result agrees with theirs.
On the other hand, the formula derived in \cite{eric} for an arbitrary
nucleus differs from ours (\ref{d}) by the absence of the magnetic
polarizability $\bb(0)$ only. The corresponding estimate presented 
in~\cite{pach} differs from
our result by the factor at $\ba_p(0)$ (2 instead of 5) and by the
absence of $\bb_p(0)$. Besides, the authors of \cite{pach}
(and of~\cite{star}) choose the inverse nucleon radius, instead of
the excitation energy, for the logarithmic cut-off in the
corresponding formulae.

The experimental data on the proton
polarizabilities, which follow from the Compton scattering, can be
summarized as follows \cite{mac}:
\[ \ba_p(0)+\bb_p(0)=(14.2 \pm 0.5)\times 10^{-4}\; \mbox{fm}^3; \]
\beq\label{co}
\ba_p(0)-\bb_p(0)=(10.0 \pm 1.5 \pm 0.9 )\times 10^{-4}\; \mbox{fm}^3.
\eeq 
Now, 
\beq\label{co1}
5\,\ba_p(0)-\bb_p(0)\,=\,2\,[\,\ba_p(0)+\bb_p(0)\,]\,
+\,3\,[\,\ba_p(0)-\bb_p(0)\,]
=(58.4 \pm 5.3)\times 10^{-4}\; \mbox{fm}^3.
\eeq 
The errors are added in quadratures.

Finally, at $\bar E_p \sim 300$ MeV the proton polarizability correction 
to the hydrogen $1S$ state is
\beq\label{res}
-\,71\,\pm\,11\,\pm\,7\; \mbox{Hz}.
\eeq
Here the first error is that of the logarithmic approximation, which
we estimate as 15\%. The second one originates from the values of the 
polarizabilities.

\bigskip

{\bf 2.}
Though being calculated rather accurately from the theoretical
point of view, the correction (\ref{res}) to the hydrogen Lamb shift
is too small to be detected experimentally. 
However, the corresponding effect
in the H-D 1S-2S isotope shift is comparable with
the experimental accuracy (150 Hz) attained for it \cite{err}.
This effect is comparable also
with the theoretical precision (70 Hz) for the contribution of the
deuteron polarizability due to relative motion of the proton and
neutron to the deuterium Lamb shift~\cite{fri1}.

The deuteron is a weakly bound system. Then it is natural
to assume that deuteron polarizability is the sum of the
polarizability due to relative motion of the nucleons and the
internal polarizabilities of the nucleons.
Simple physical arguments, supported by model estimates,
demonstrate that nucleon polarizabilities in deuteron coincide
with polarizabilities for free nucleons, well within the accuracy of
our logarithmic approximation. Therefore, in the corresponding effect
in the H-D isotope shift the proton contributions
cancel, and we are left with that of a neutron (with opposite sign)
\begin{equation}\label{hd}
\delta V_{H-D}=\,\alpha m_e\,(\,5{\alpha}_n(0)-{\beta}_n(0)\,)\;
                                \ln\frac{\bar E_n}{m_e}\;\delta({\bf r}).
\end{equation}
The neutron electric polarizability is \cite{neut}
\[ \alpha_n(0)=(9.8 \; {}^{ +1.9}_{ -2.3} )
                             \times 10^{-4}\; \mbox{fm}^3; \]
Its magnetic polarizability $\beta_n$ is not known.
Under the assumption that $\beta_n$ does not change the result 
considerably, this contribution to the difference between the Lamb
shifts of the ground states of hydrogen and deuterium is
\beq\
61 \pm \,10 \pm \, 12 \mbox{Hz}.
\eeq

The corresponding contribution to the isotope shift between hydrogen
and deuterium 1S-2S transition
due to the internal polarizabilities of the nucleons can be estimated as
\beq\label{a}
53 \pm \,9 \pm \, 11 \mbox{Hz}.
\eeq
 
\vspace*{.5cm}

We are grateful to A.I. Milstein and A.A. Pomeransky for useful
discussions. We wish to thank R.N. Faustov, S.B. Gerasimov, and J.L. Friar
for bringing to our attention papers \cite{star,eric}.
We are grateful to P.J. Mohr and B.N. Taylor for their advice to publish
this result, and to S.A. Coon for pointing out a misprint in the original
version. 
We acknowledge the support by the Russian Foundation for Basic
Research through grant No. 98-02-17797, Federal Program Integration --
1998 through Project No. 274.

\end{document}